\documentclass[11pt,a4paper]{article}
\usepackage[utf8]{inputenc}
\usepackage[english]{babel}
\usepackage{amsmath}
\usepackage{amsfonts}
\usepackage{amssymb}
\usepackage{graphicx}
\usepackage{listings}
\usepackage{float}
\usepackage{booktabs}
\usepackage{geometry}
\usepackage{subfig}
\usepackage{cite}

\usepackage{vmargin}

\setpapersize{A4}
\setmargins{3cm}       % margen izquierdo
{1.5cm}                        % margen superior
{15.25cm}                      % anchura del texto
{23.42cm}                    % altura del texto
{10pt}                           % altura de los encabezados
{1cm}                           % espacio entre el texto y los encabezados
{0pt}                             % altura del pie de página
{2cm}                           % espacio entre el texto y el pie de página

\usepackage{graphicx}
\title{Design of slow-light-enhanced bimodal interferometers using dimensionality reduction techniques}
\author{\emph{Luis Torrijos-Morán, and Jaime García-Rupérez}\\
\vspace{0.75cm}
\small{Nanophotonics Technology Center}\\
\small{Universitat Politècnica de València, 46022 Valencia, Spain.}\\
\small{luitorm2@ntc.upv.es}} 

\makeatletter         
\def\@maketitle{
\begin{center}
{\Huge \bfseries \sffamily \@title }\\[4ex] 
{\Large  \@author}\\[4ex] 
\end{center}}
\makeatother

\begin{document}
\maketitle
\sloppy
% \thispagestyle{empty}
%\newpage

%\newpage
%\vspace{2.5cm}

\vspace{-0.25cm}

\begin{changemargin}{0.6cm}{0.6cm} 
Interferometers usually require long paths for the ever-increasing requirements of high-performance operation, which hinders the miniaturization and integration of photonic circuits into very compact devices. Slow-light based interferometers provide interesting advantages in terms of both compactness and sensitivity, although their optimization is computationally costly and inefficient, due to the large number of parameters to be simultaneously designed. Here we propose the design of slow-light-enhanced bimodal interferometers by using principal component analysis to reduce the high-dimensional design space. A low-dimensional hyperplane containing all optimized designs is provided and investigated for changes in the silicon core and cladding refractive index. As a result, all-dielectric single-channel interferometers as modulators of only 33 $\mu$m\textsuperscript{2} footprint  and sensors with 19.2×10\textsuperscript{3} 2$\pi$rad/RIU·cm sensitivity values are reported and validated by two different simulation methods. This work allows the design and optimization of slow light interferometers for different applications by considering several performance criteria, which can be extended to other photonic structures.
\end{changemargin}

\section{Introduction}

Optical interferometers are essential components in today integrated photonics. Among others, they are widely used in modulators, switches, sensors and programmable photonic circuits \cite{Reed2010,Kumar2013,Kozma2014,Bogaerts2020}, due to their ability of converting a change in the refractive index (RI) into a shift of the relative phase between two propagating modes. To enhance the phase shift, further optimization either in the material or in the geometry is required to maximize light-matter interaction. In sensing interferometers, a strong field interaction with the cladding is desired to increase the sensitivity \cite{Prieto2003}, for instance by employing plasmonic configurations \cite{Bozhevolnyi2006,Gao2011} or lower RI contrast materials such as silicon nitride \cite{Munoz2017}. In contrast, typical electro-optical modulators based on Mach-Zehnder (MZI) interferometers in silicon-on-insulator (SOI) platforms require from highly confined modes within the waveguide core to assure a complete overlap of the optical and the electric field \cite{Liu2004,Liao2005,Green2007}. Similarly, SOI thermo-optic switches are also designed to support highly confined   optical modes to optimize the interaction of the optical field with the core RI changes due to the thermo-optic coefficient of silicon \cite{Geis2004a,Sun2010,Watts2013}. Concurrently, interferometers performance can also be improved by engineering new structures and including them in one of the MZI arms, as in the case of slot waveguides \cite{Liu2013a,Sun2015} or subwavelength grating (SWG) structures \cite{Bock2010,Sumi2017}, although roughness scattering losses must be considered compared to strip waveguides \cite{Ita2018}.

On the other hand, the physical path of the interferometer limits the size of the device as long interferometric structures are typically designed to provide high-performance operation, which hinders the fabrication of densely integrated optical circuits \cite{Sorger2012}. To solve this issue, long MZIs are proposed as compact interferometers by including low-loss compact bends with small radius in the MZI arms \cite{Densmore2009}, as well as bimodal (BiM) waveguide sensors where the interferometry is carried out in a single-channel structure \cite{Zinoviev2011}. More sophisticated BiM waveguides based on SWG structures have also been reported as compact and high-sensitivity sensors without the need of additional photonic structures \cite{Torrijos-Moran2019,Torrijos-Moran2019a}. Another interesting approach is to increase the optical path of the interferometer using slow light structures \cite{Shaw1999,Soljacic2002}, while maintaining a reduced physical length. This is the case of photonic crystals (PhCs), either combining them in the arms of a MZI to develop modulators \cite{Jiang2005,OFaolain2010a,Brimont2011a} and sensors \cite{Qin2016}, or by fully integrating an interferometric scheme in a 2D hole-patterned PhC for switching purposes \cite{Camargo2004,Nakamura2004}. In this scenario, further improvement was made by including a bimodal behavior in a single-channel slow light waveguide \cite{Torrijos-Moran2021,Torrijos-Moran2021b}, demonstrating high-performance interferometers as modulators, switches, and sensors, with extremely reduced footprints compared to the abovementioned approaches. However, the optimization of these types of periodic structures is not straightforward because of the multiple design dimensions to consider. Several optimization methods based on genetic algorithms \cite{Hakansson2005,Covey2013}, gradient-based optimizations \cite{Jensen2004,Niederberger2014} or particle swarm \cite{Ma2013,Watanabe2017} are proposed as useful tools to design high-performance grating couplers, SWG structures or PhCs, among others. More recently, machine learning techniques have also been demonstrated for designing optimized nanophotonic components using artificial networks \cite{Peurifoy2017,Turduev2018}. Nevertheless, all these optimization methods focus on optimizing a single criterion, which makes it very difficult to obtain a general perspective of the overall device performance. In contrast, dimensionality reduction techniques offer interesting design tools to optimize multi-parameter structures by taking into consideration different performance criteria \cite{Melati2019}, which eases the design process for the desired application.

In this work, we propose the design and optimization of multi-variable slow-light-enhanced bimodal interferometers by using principal component analysis (PCA). A figure of merit (FoM) is provided, which allows us to characterize the desired bimodal band structure and extrapolate it to other different PhC designs. By using PCA of the FoM, we explore and optimize the low-dimensional design space for changes in both the silicon core and the cladding RI. Three different interferometers are presented and optimized as sensors and modulators, offering significant improvements both in sensitivity  and compactness compared to other existing configurations.

\section{Figure of merit definition}
In a PhC, when two bands of the same polarization and parity intersect, they repeal each other and form an anti-crossing point  where electromagnetic wave propagation is forbidden \cite{Notomi2001}. Furthermore, it has already been demonstrated how slow light bimodal interferometers, based on two modes with a large group index difference, can be designed in the vicinity of the anti-crossing point \cite{Torrijos-Moran2021}. Figure 1a schematically shows the first three bands of a band structure in a PhC formed by the fundamental mode folded into the first Brillouin zone and the higher order mode. Figure 1b and c depict the third bimodal band and its corresponding group index for both modes as a function of wavelength. The group index is defined as $n_g=c/v_g$ where $c$ is the speed of light in the vacuum and $v_g$ the group velocity, which is related to the slope of the wavelength versus the wavevector. Note that the maximum group index difference is obtained at $\lambda_{1}$, where the higher order mode reaches the end of the first Brillouin zone and becomes slow light. Conversely, at $\lambda_{2}$, both modes present a slow light behavior and almost no group index difference is observed. In this work, we define a FoM to characterize the curvature of the third bimodal band  in order to quantify the group index difference and the bandwidth of the bimodal region. The FoM is mathematically expressed as

\begin{equation}
FoM=\int_{\lambda_{1}}^{\lambda_{2}} (n_{g2} - n_{g1}) \,d\lambda
\end{equation}

\noindent
where $\lambda_{1,2}$ are the entire wavelength limits   of the bimodal region, and $n_{g2,1}$ are the group index of the  higher order and fundamental mode, respectively. Therefore, the FoM may be defined as the area under the curve of group index difference as a function of wavelength. By optimizing this variable, we assure a bimodal region with large group index contrast and bandwidth, which are the main parameters in order to maximize the optical path of the interferometer. The higher the group index difference, the larger the phase shift obtained for a given change in the RI, regardless the physical length. Moreover, the FoM provides information about the bimodal band curvature, which is common to other PhC designs in which an anti-crossing point is formed, see Fig . 1a. This fact will allow us to compare between PhC configurations of the same design, and between different designs with new geometric shapes, and thus, with a different light-matter interaction.

\begin{figure}[hb!]
\centering\includegraphics[scale=0.65]{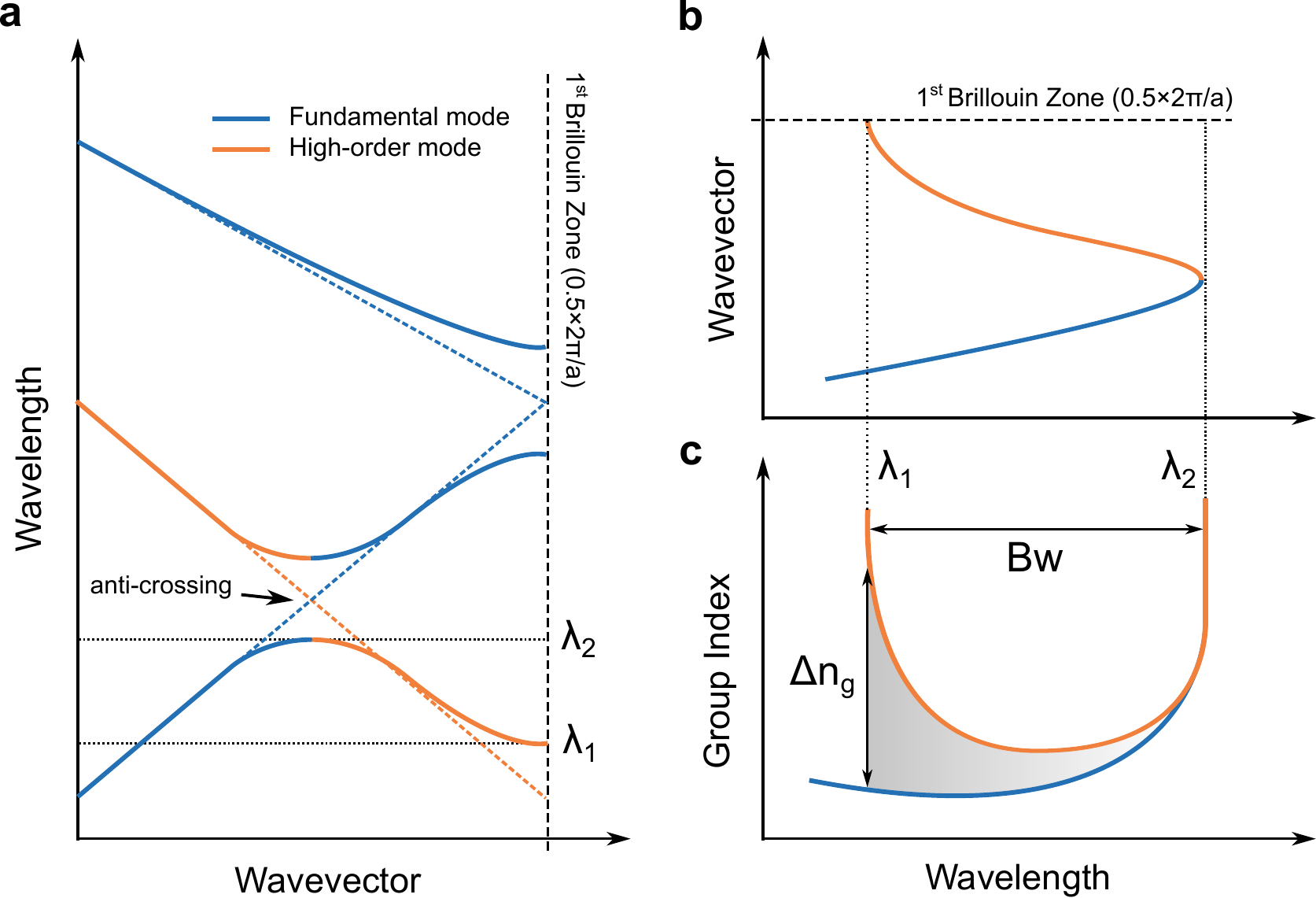}
\caption{Bimodal band structure representation. a) The first three bands are shown as well as the contribution of the fundamental and high-order modes.  The anti-crossing point of a PhC is also depicted in the intersection between the second and third band. Dashed lines represent the hypothetical band structure of a homogeneous waveguide without periodicity. b) Third band considered to perform the interferometry and c) group index as a function of wavelength. The gray shaded area represents the FoM.}
\end{figure}

\section{Design optimization}

In the following section, three designs based on one-dimensional PhC waveguides made of silicon over silica cladding are investigated for sensing and modulating purposes. The unit cell geometry is engineered to obtain a bimodal band structure with a similar curvature than the third band presented in Fig. 1b, so that it can be quantified by the FoM previously defined. To this end, a first exploration of the multi-parameter design space is carried out by using MIT photonics band (MPB) free software \cite{Johnson2001}. MPB computes definite-frequency eigenstates of Maxwell’s equations by using plain wave expansion numerical methods, to obtain the band structure of the PhC. The first three bands for the transverse electric (TE)-like polarization have been computed. An initial low-resolution sweep of $n^D$ simulations is carried out, where $n$ is the number of values for each dimension and $D$ the number of design dimensions. A collection of good designs providing values of FoM above a certain threshold is selected from this initial sweep. These designs are subsequently used in the PCA to extract the sub-dimensional design space \cite{Melati2019}. PCA finds among the multi-dimensional design space a set of orthogonal vectors that maximizes the variance of the FoM. Specifically, a two-dimensional hyperplane formed by the first two principal components is selected, which critically reduces the design space dimensionality and enables a low-computational exhaustive mapping. Thereby, a design $k$ with dimensions $\textbf{L}_k=[L_{1,k}...L_{D,k}]$ can be expressed as

\begin{equation}
\textbf{L}_k = \alpha_k \textbf{V}_{1 \alpha \beta} + \beta_k \textbf{V}_{2 \alpha \beta} + \textbf{C}_{\alpha \beta}
\end{equation}

\noindent
where $\alpha$, $\beta$ are the coefficients that describe any design in the hyperplane formed by the first two principal components vectors $\textbf{V}_{1,2}$, and $\textbf{C}_{\alpha \beta}$ is the origin constant vector. High-resolution sweeps of the resulting 2D hyperplane are now reachable, with a wide collection of good designs for different configurations. Besides, an exhaustive mapping of the hyperplane can be further evaluated for different criteria beyond the FoM. More precisely, we are going to investigate the interferometer as sensors and modulators, thus for changes in the cladding and silicon RI, respectively, and also for fabrication robustness.

The first 1D PhC design considered is shown in Fig. 2a, based on a periodic corrugated waveguide in the propagation direction. This design has already been investigated in Ref. \cite{Torrijos-Moran2021}, but here we intend to optimize it by using PCA. The unit cell dimensions are shown on the right-side part of Fig. 2a, for the periodicity $a$, central waveguide width $w$, transversal element length $w_e$, width $w_i$ and height $h$. Note that a rectangular taper for the efficient bimodal excitation is placed between the single mode in-output waveguide of width $w_s$ and the periodic bimodal structure, although these parameters do not affect the band calculation. The height is set to 220 nm, a standard in SOI wafers, so that a four-dimensional design of the unit cell is obtained $(a,w,w_e,w_i)$. To reduce the computational resources, the number of values for each dimension $n$ is set to a low value of 4. Thus, an initial sweep of 44 = 256 simulations has been computed to obtain the band diagram in MPB, and its respective FoM, of the entire multi-dimensional space. An initial collection of 44 good designs with a FoM over a threshold of 270  has been selected for the PCA. This value has been selected to obtain a relatively large number of designs in the initial collection.

\begin{figure}[t!]
\centering\includegraphics[scale=0.65]{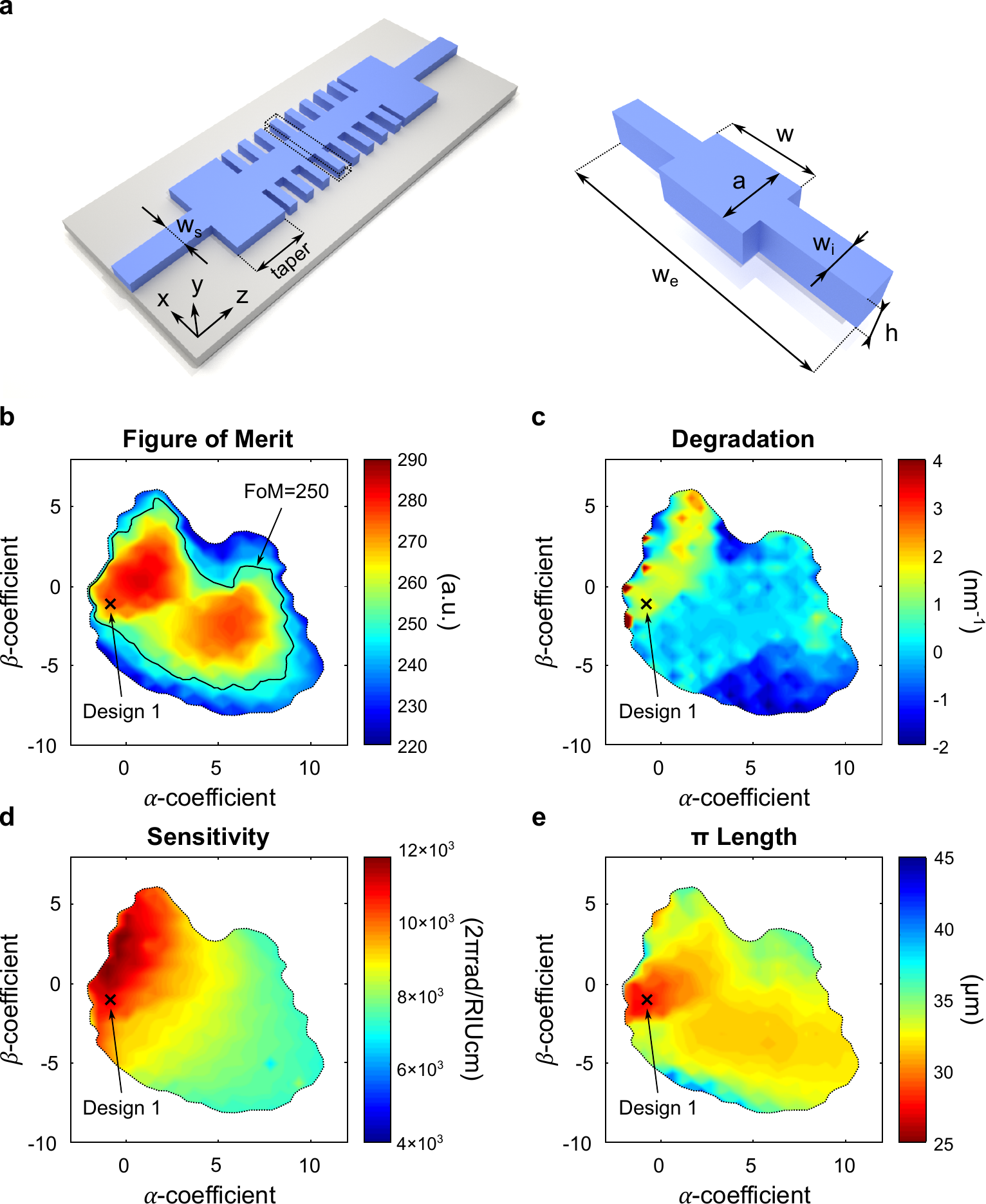}
\caption{Corrugated waveguide design optimization. (a) Sketch of the complete slow- light-enhanced interferometer based on a bimodal PhC waveguide, and two single-mode waveguides at the input and output. The unit cell dimensions are also specified on the right-hand side of the figure. (b) Colormap of the figure of merit PCA, for the first and second   principal components of the collection of good designs. (c) Degradation of the F igure of merit for a width variation of +-10 nm in all the specified dimensions except for the period. (d) Upper cladding RI sensitivity normalized to a 1 cm length at $\lambda_0$ wavelength. (e) Length required for a $\pi$ phase shift at $\lambda_0$ wavelength and a given silicon RI increment of 0.01.}
\end{figure}

Figure 2b depicts the exhaustive mapping, as a function of $\alpha$ and $\beta$ coefficients, of the 2D hyperplane after the PCA with a high-resolution sweep of 30×30 MPB simulations. The black solid contour represents those designs with a FoM of 250, which is the value calculated in Ref. \cite{Torrijos-Moran2021}. Those FoM values out of the color bar range are not represented to emphasize the optimized values.   A relatively large region of different designs with optimized FoM values is obtained, within which two optimized FoM areas with values around 280 are observed. Likewise, the hyperplane  of optimized designs is explored for small fabrications deviations of +- 10 nm in the design dimensions. The degradation of the FoM is shown in Fig. 2c, calculated as $d=-(F^+ + F^- -2F_0)/(2\Delta \delta_w)$) where $F^{+,-}$ is the FoM for a positive and negative deviation, respectively, $F_0$ the initial FoM without deviations and $\Delta \delta_w$ is the absolute deviation width considered. Note that a positive degradation means a reduction of the FoM when deviations occur, whereas a negative degradation will be translated into an increment of the FoM.

Figure 2d and e show the colormap for the sensitivity and $\pi$ length, which are strongly related with the light-matter interaction of the propagating modes with the cladding and silicon core, respectively.  The phase shift has been calculated for a given change in the cladding and in the silicon RI. Due to the dispersion of these interferometers, the phase shift strongly depends on the operating wavelength, being much higher for those regions near $\lambda_1$ in Fig. 1a. However, coupling losses are also incremented within the slow light region, so that a trade-off between sensitivity and losses is obtained \cite{Hugonin2007}. Due to this fact, the phase shift is calculated at $\lambda_0$ wavelength where the group index of the higher order mode is limited to 20, since this value has been experimentally demonstrated in similar structures \cite{Torrijos-Moran2021}. At this wavelength, the sensitivity to cladding RI changes shown in Fig. 2d and normalized to a length of 1 cm, has been calculated as $S=\Delta \phi / \Delta n_c$, where $\Delta \phi$ is the phase shift difference for a cladding RI change $\Delta n_c$ of 0.01. Similarly, changes in the silicon RI have also been investigated by calculating the length required to obtain a phase shift of $\pi$, see Fig. 2e. As in the previous case, it has been obtained at $\lambda_0$ wavelength as $L_\pi=\pi/\Delta\beta$ where $\Delta\beta$ is the increment in the propagation constants of both modes for a given change in the silicon RI of 0.01. Design 1, marked with a cross in Fig. 2, has been chosen from the hyperplane at the region where both the sensitivity and the $\pi$ length reaches its maximum, and where considerable low values of degradation are obtained. A FoM of 272.2 with a degradation of 1.41 nm\textsuperscript{-1}, a sensitivity of 10,870 2$\pi$rad and a $\pi$ length of 27.5 $\mu$m has been obtained for the corrugated waveguide optimization of Design 1.

\begin{figure}[b!]
\centering\includegraphics[scale=0.65]{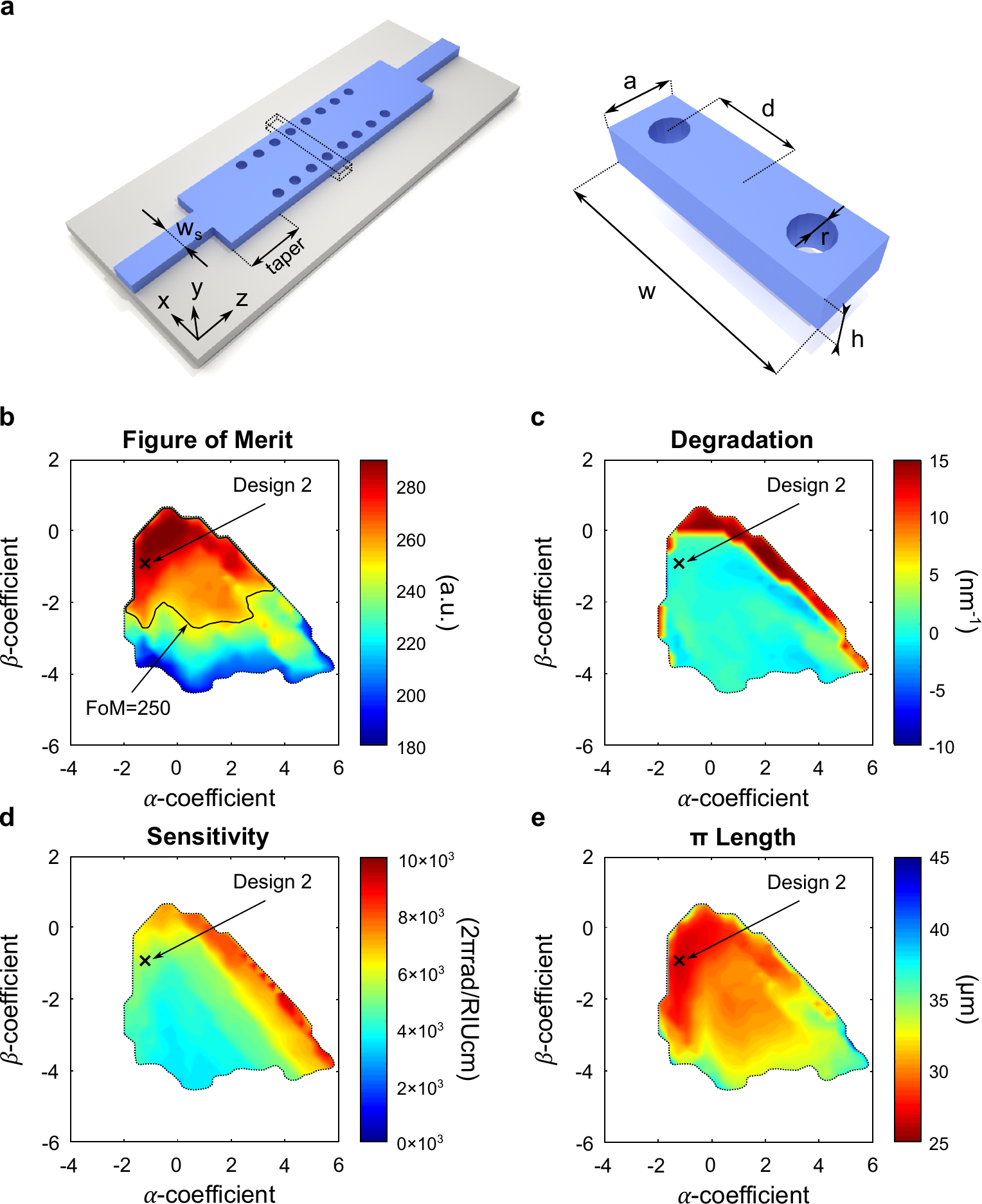}
\caption{Double hole waveguide design optimization. (a) Sketch of the complete slow- light-enhanced interferometer with the unit cell dimensions of the 1D PhC. Colormap of the two principal components of the (b) figure of merit, (c) degradation for deviations of +-10 nm, (d) normalized sensitivity to 1 cm and (e) and $\pi$ length for a silicon RI increment of 0.01.}
\end{figure}

\begin{figure}[b!]
\centering\includegraphics[scale=0.65]{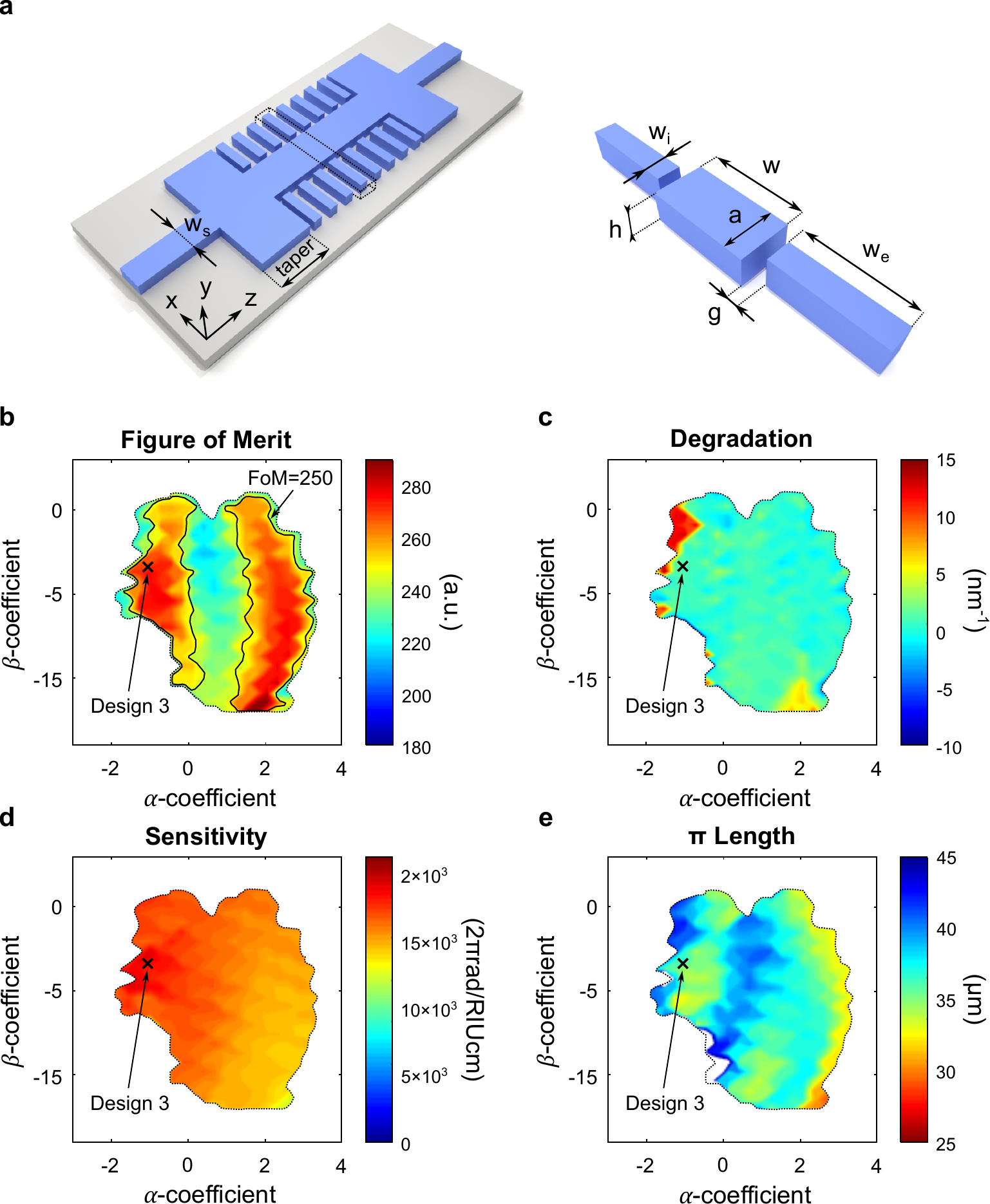}
\caption{Multi box waveguide design optimization. (a) Sketch of the complete slow- light-enhanced interferometer with the unit cell dimensions of the 1D PhC. Colormap of the two principal components of the (b) figure of merit, (c) degradation for deviations of +-10 nm, (d) normalized sensitivity to 1 cm and (e) and $\pi$ length for a silicon RI increment of 0.01.}
\end{figure}

The second design is presented in Fig. 3a, consisting of a double array of circular holes over a uniform rectangular waveguide, forming a 1D PhC in the propagation direction. The dimensions of the unit cell are shown in the right-hand side of Fig. 3a, for the periodicity $a$, waveguide width $w$, distance from the waveguide center to the holes center $d$, hole radius $r$ and height $h$. As in the previous case, the height has been fixed to 220 nm, so that a four-dimensional design space is obtained $(a,w,d,r)$. An initial low-resolution sweep of 256 simulations has been carried out to make a first exploration. In this case, 23 good designs with a FoM above a threshold of 280   are selected for the PCA. Figure 3b shows the FoM colormap as a function of $|alpha$ and $\beta$ coefficients. A contour delimiting those good designs with a FoM value higher than 250 is depicted, which demonstrates that this 1D PhC design can also be improved in comparison to \cite{Torrijos-Moran2021}. It should be also noted that now the maximized area of the FoM is localized on a certain region of the colormap, close to the edge of good designs. Figure 3c shows that these designs present high degradation values due to the deterioration of the band curvature at these regions where FoM is drastically deteriorated. Figure 3d and e show the sensitivity and $\pi$ length, respectively, both calculated at $\lambda_0$ wavelength. High-sensitivity designs are obtained near the right-side edge of the hyperplane where high-degradation values were also obtained, while the $\pi$ length is clearly optimized in the opposite left side part of the colormap. Design 2 marked with a cross in Fig. 3 has been chosen from a region of optimized $\pi$ length designs and low degradation. Specifically, a FoM of 284.6 with a degradation of 0.28 nm\textsuperscript{-1}, a sensitivity of 5,364 2$\pi$rad and a $\pi$ length of 25.9$\mu$m has been obtained for the double hole waveguide optimization of Design 2.

Finally, the third 1D PhC design is presented in Fig. 4a and consists of two independent transversal elements periodically placed besides a central uniform waveguide, forming a kind of multi-box periodic waveguide in the propagation direction. In this case, a five-dimensional design space is obtained since a small gap is defined within the structure. By using PCA, more complex designs can also be optimized which enables new geometric shapes with new interesting light matter interaction. The unit cell dimensions are shown for the periodicity $a$, central waveguide width $w$, transversal element length $w_e$, width $w_i$, gap between the central waveguide and the transversal element $g$ and height $h$. The height is fixed to 220 nm as in previous designs, although in this case a five-dimensional design is obtained. An initial sweep of 35 = 243 simulations is computed to select 42 good designs with a FoM above 260. Figure 4b shows the FoM colormap as a function of the first and second principal components  where two main regions of designs with a FoM value higher than 250   are obtained. The degradation is equally distributed, see Fig. 4c, while the sensitivity in Fig. 4d is homogeneously distributed and maximized for those designs in the left side part of the FoM colormap. Nevertheless, low-performance regarding the $\pi$ length is obtained for the entire PCA mapping in comparison with the previous design, see Fig. 4e. Therefore, Design 3, marked with a cross in Fig. 4, has been chosen in the region of maximized sensitivity with a FoM of 275.7, degradation of 1.09 nm\textsuperscript{-1}, sensitivity of 19,196 2$\pi$rad and a $\pi$ length of 36.6 $\mu$m.

\section{Comparison of the results and discussion}
The resulting optimized designs from the PCA are presented in Table 1, which shows a comparison between the main results regarding the FoM, fabrication deviations and changes in the RI of the cladding and the core of the structure. Dimensions are sorted in the appearance order in which they have been previously detailed. Note that all designs present a higher FoM value than the one reported in Ref. \cite{Torrijos-Moran2021}, which demonstrates an enhancement of the optical path in slow-light-enhanced bimodal waveguides by using PCA. Moreover, further improvement in terms of both sensitivity and $\pi$ length is presented. Compared to \cite{Torrijos-Moran2021}, a specific 22 \% reduction in the physical length of the interferometer as a modulator (changes in the silicon RI) is obtained for Design 2, whereas the sensitivity is increased by a factor of 1.85 for Design 3, acting as a sensor (changes in the cladding RI). By reducing the design space dimensionality, we develop a parallel optimization for both the optical path and the light matter interaction of the interferometer. Furthermore, this method has been employed to come up with new geometrical designs with very different performance, as for example the sensitivity contrast between Designs 2 and 3.

\begin{table}[t!]
\caption{\textbf{PCA Optimized designs comparison}}
\scalebox{0.995}{%
\begin{tabular}{@{}ccccccc@{}}
\toprule
Design    & {}$\alpha$, $\beta${}         & Dimensions \small(nm)          & FoM   & Deg. \small(nm\textsuperscript{-1}) & S \small(2$\pi$rad) & L\textsubscript{$\pi$} \small($\mu$m) \\ \midrule
Ref. \cite{Torrijos-Moran2021}      & None              & {[}370,600,1400,220{]}   & 249.8 & 0.62        & 10,337           & 33       \\
1 (corr.) & {[}-0.93,-1.31{]} & {[}378,560,1638,171{]}   & 272.2 & 1.41        & 10,870           & 27.5     \\
2 (hol.)  & {[}-1.24,-1.03{]} & {[}387,1319,439,100{]}   & 284.6 & 0.28        & 5,364            & 25.6     \\
3 ( box.) & {[}-0.96,-3.17{]} & {[}360,726,722,168,90{]} & 275.7 & 1.09        & 19,196           & 36.6     \\ \bottomrule
\end{tabular}}
\end{table}

\begin{figure}[b!]
\centering\includegraphics[scale=0.65]{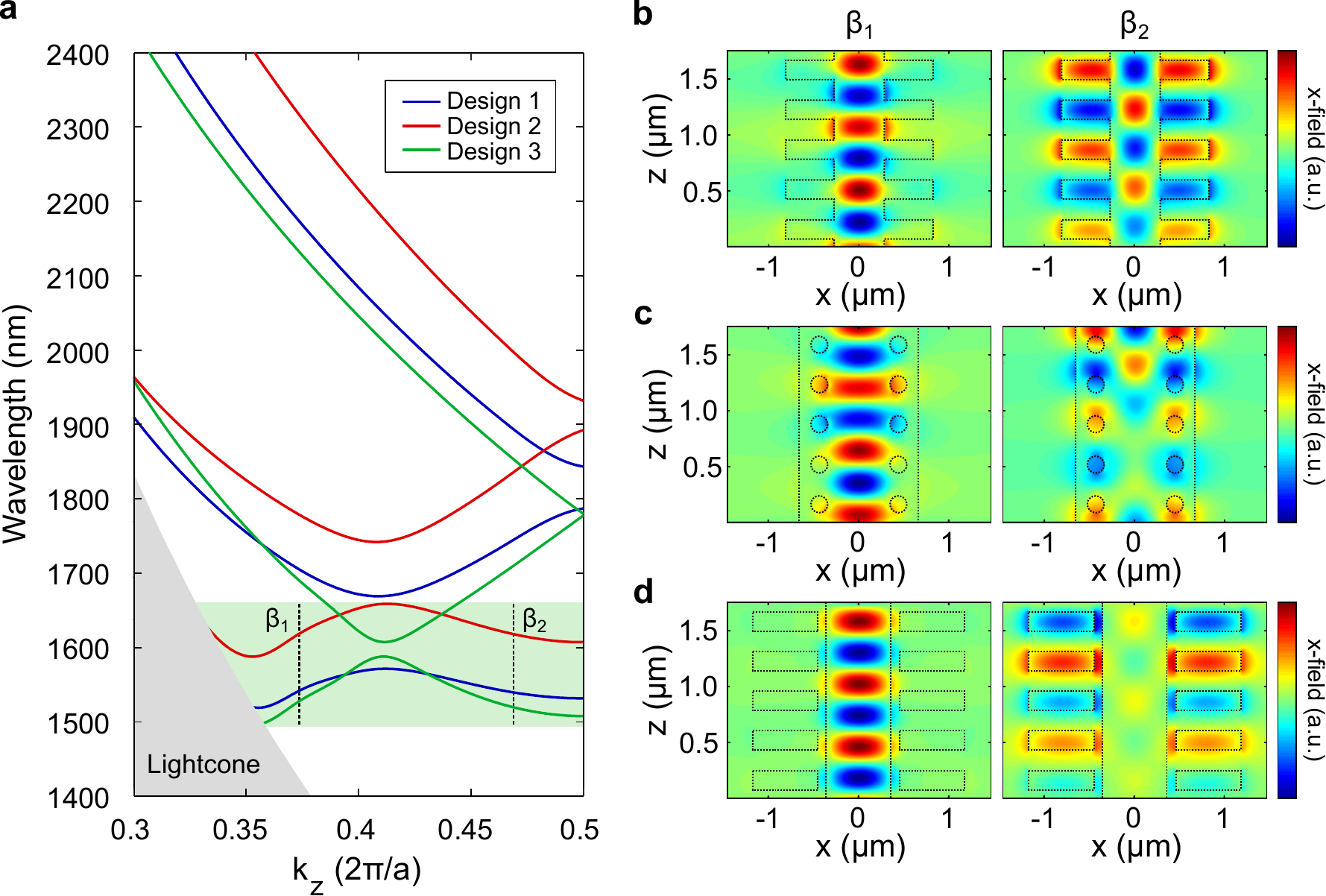}
\caption{(a) First three TE-like polarized bands of the band diagram for the optimized designs in the PCA. The green shaded area represents the bimodal region. (b) Real part of the $x-$component field in the corrugated waveguide design for the fundamental mode at $k_z=0.375$ 2$\pi / a$ and the higher order mode at $k_z=0.475$ 2$\pi / a$, left and right colormaps, respectively. (c) Field distribution in the double hole waveguide and (d) multibox waveguide designs. The dashed line represents the silicon structure shape.}
\end{figure}

Figure 5a shows the first three bands for the TE-like polarization for the three designs presented in Table 1. The region of interest is located around 1550 nm at the third band curvature, marked with the green shaded area of Fig. 5a. As it was schematically explained in Fig. 1, in this region an anti-crossing point is formed, below which the bimodal desired operation is obtained, see $\beta_{1,2}$ in Fig. 5a. The band structure formation for all designs is very similar, specifically the curvature of the third band, in accordance with the FoM results obtained in Table 1. Figure 5b, c and d show the field distribution for Designs 1, 2 and 3, respectively. Both the fundamental ($\beta_1$) and the higher order ($\beta_2$) modes are shown. In all these cases, the fundamental mode is highly confined within the center of the waveguide, while the higher order mode is partially localized in the periodic pattern of the structure. Note that in the right graph of Fig. 5c, the $\beta_2$ mode is more confined within the silicon structure than in the case of the right graph of Fig. 5d, where a high part of the field is localized within the gap. This fact explains why Design 2 is highly sensitive for changes in the silicon RI and Design 3 for changes in the cladding RI, and thus their difference between the sensitivity and $\pi$ length results.

For the sake of completeness, the results have also been validated with 3D finite difference time domain (FDTD) simulations using the CST Studio software. In this case, the unit cells are periodically disposed in the propagation direction with its correspondent taper configuration of \textasciitilde 1.5 $\mu$m length and accessed in and out with a single mode waveguide of 450 nm width. The fundamental TE mode of the input waveguide excites both modes in the bimodal PhC, which similarly contribute to the excitation of the fundamental TE mode at the output waveguide. Thus, the bimodal interference pattern can be observed in the transmitted spectra. To measure the phase shift, we need to calculate the relationship between the wavelength shift of a constructive  interference in the spectrum and the free spectral range (FSR) by using the following expression: $\Delta \phi = \Delta \lambda / FSR$. Figure 6a and b depict the phase shift of the lowest wavelength interference for all designs as a function of linear changes in the cladding and silicon RI, respectively.

\begin{figure}[b!]
\centering\includegraphics[scale=0.70]{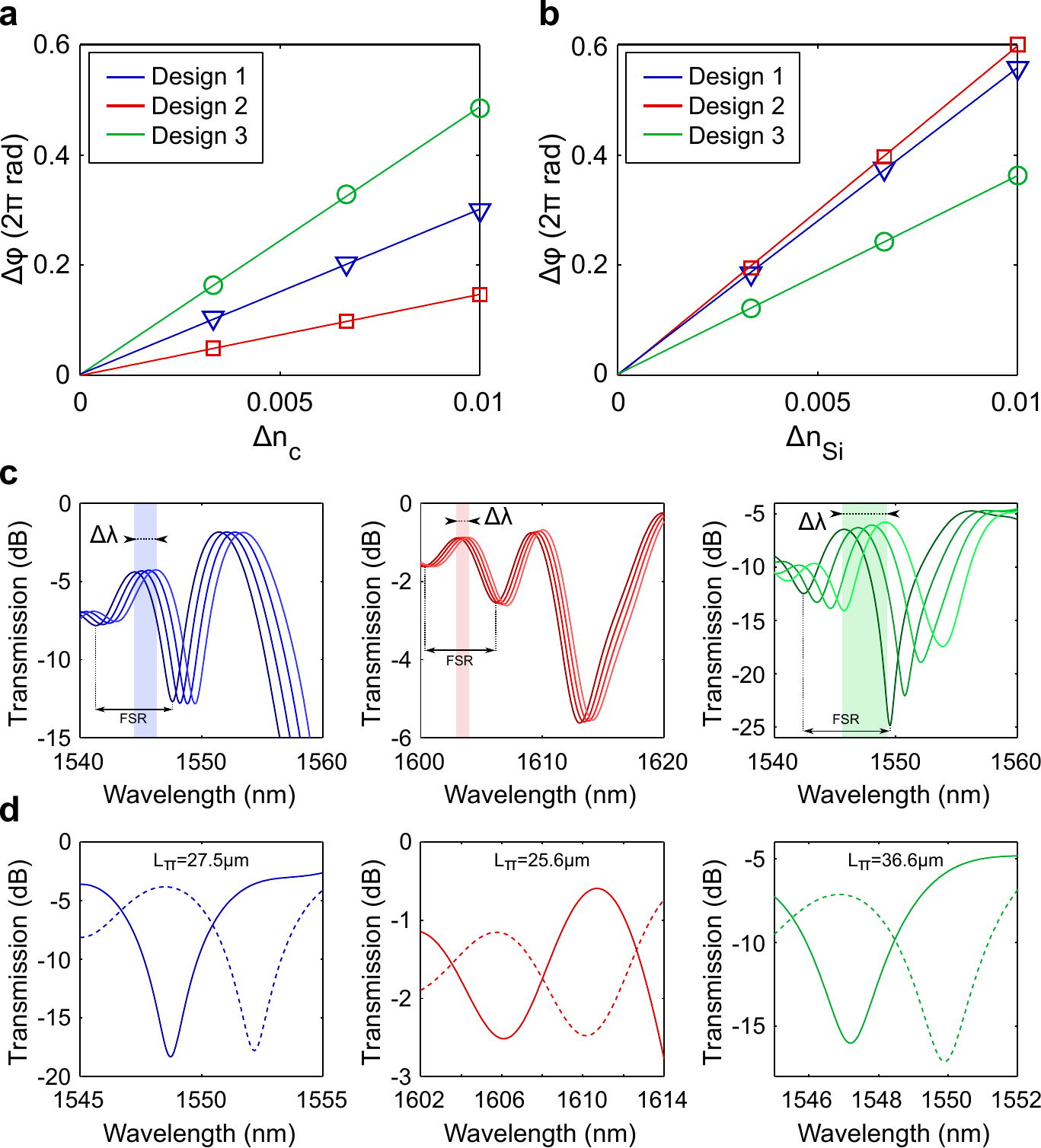}
\caption{Comparison of the optimized designs in FDTD simulations. Phase shift of the first constructive interference for (a) changes in cladding and (b) in the silicon refractive index. (c) Transmission spectra of the slow light bimodal interferometers with N=100 periods and for different cladding RI (a linear sweep from 1.444 to 1.454). All PCA optimized configurations are shown, Design 1 (blue colored left graph), Design 2 (red colored middle graph of Fig. 6c), and Design 3 (green colored right graph). (d) Transmission spectra for all PCA optimized designs with a length $L_{\pi}$. Solid lines represents the spectra for a RI index of 3.4777 and dashed line for 3.4877.}
\end{figure}

Note that the most sensitive configuration to cladding changes in Design 3, which is in a good agreement with the results shown in Table 1. Likewise, Design 2 is the most sensitive to silicon RI variations, hence the lower $\pi$ length is obtained for this configuration. Design 1 presents a balanced behavior for changes in both the cladding and core, which also matches with previous results. Figure 6b shows the considered interference in the transmitted spectra for all designs under cladding RI changes. The FSR is almost equal in all of them, while the wavelength shift is clearly higher for Design 3 (green right-side graph), which explains the results of Fig. 6a. Similarly, Fig. 6d shows the spectra for a change in the silicon core RI of 0.01 and for 1D PhCs with the $\pi$ lengths calculated in the PCA. The FDTD results perfectly match previous MPB simulations, an a clear $\pi$ phase shift is observed in the spectra.

\section{Conclusions}
We have demonstrated a method to design and optimize slow-light-enhanced bimodal interferometers by using dimensionality reduction techniques. A FoM has been introduced to characterize the bimodal band curvature of the 1D PhC and PCA has been employed to optimize the optical path of the interferometers in terms of group index and bandwidth. A low-dimensional 2D hyperplane containing the optimized designs has been obtained, which allows us to explore different performance criteria as the degradation to fabrication deviations, bulk sensitivity and $\pi$ length for changes in the silicon RI. As a result, three different single-channel 1D PhC interferometers have been designed, with remarkable improvements regarding other similar slow-light-enhanced bimodal structures \cite{Torrijos-Moran2021}. Specifically, simulations show that the physical interferometer modulation length has been reduced by 22 \% for silicon RI variations, which means that this device can be integrated in an all-dielectric structure of only 33 $\mu$m\textsuperscript{2} footprint. In comparison with the literature, these results correspond to a reduction of more than two orders of magnitude respect to conventional MZI structures [10] and more than one order compared to slow-light based interferometers [26, 29]. Moreover, a sensor with values of 19.2 ×10\textsuperscript{3} 2$\pi$rad/RIU·cm has been reported, which means nearly twice as high as cited slow-light-enhanced bimodal interferometers [33], and more than one order of magnitude compared to standard MZI sensors and bimodal waveguides [5, 22]. Overall, these findings open up new avenues to design and optimize periodically structured devices which may lead to novel applications in different  areas with enhanced properties.

\subsection*{Fundings}

The authors acknowledge funding from the Generalitat Valenciana through the AVANTI/2019/123, ACIF/2019/009 and PPC/2020/037 grants, from the Spanish Government through the PID2019-106965RB-C21-PHOLOW project and from the European Union through the operational program of the European Regional Development Fund (FEDER) of the Valencia Regional Government 2014–2020.

\subsection*{Disclosures}

The authors declare no conflicts of interest.

%%%%%%%%%% If using BibTeX:
\bibliography{Preprint}

\end{document}